\def\bi{$\rm Bi_2Sr_2CaCu_2O_8$\ }  
\def\laal{$\rm LaAlO_3$\ }     
\def\ha{$H_a$\ }     
\def\hmw{$h_1$\ }    
\def\hm{$h_m$\ }     
\def\tc{T$_c$\ }     
\def\hir{$H_{ir}(T)$\ }  
\def\bfi{$B_{\Phi}$\ }  
\def\ur{$\rm ^{238}U$} 

\documentclass[preprint,eqsecnum,aps]{revtex4}
\usepackage[dvips]{graphicx}
\usepackage{float}
\usepackage{amssymb}
\draft
\preprint{}

\begin{document}

\title{Shift of the surface-barrier part of the irreversibility line 
due to columnar defects in $\rm \mathbf{Bi_2Sr_2CaCu_2O_8}$\ thin films}

\author{Yu.\,Talanov}
\address{Zavoisky Physical-Technical Institute, 420029, Kazan, Russia} 
\author{H.\,Adrian, M.\,Basset, G.\,Jakob}
\address{Johannes Gutenberg-Universit\"at, Institut f\"ur Physik, D-55099 Mainz, Germany}
\author{G.\,Wirth}
\address{Gesellschaft f\"ur Schwerionenforschung, 64291 Darmstadt, Germany}

\begin{abstract}
In the present work we report the results of studying the influence of the
uranium-ion irradiation of the \bi thin films on the high-temperature part
(close to \tc) of their irreversibility line. We studied irreversible
properties of the films by measuring the hysteresis of nonresonant
microwave absorption. The results have revealed the shift of
irreversibility line towards low temperatures and magnetic fields. The
effect is most significant for the films irradiated with large doses,
$B_{\Phi}>1$\,T. This fact is in good agreement with the theoretical
prediction by Koshelev and Vinokur of suppression of surface barrier by
columnar defects.
\end{abstract}

\keywords{ vortex matter \sep columnar defects \sep bulk pinning \sep surface pinning \sep irreversibility line \sep microwave absorption \sep \bi films}

\maketitle

\section{Introduction}

It is well known that in some high-\tc superconductors intrinsic (natural)
pinning centers have rather little efficiency. They can not support 
undissipative current flow with large enough density. In crystals and films
of \bi the critical current density is very low in a wide area of magnetic
field and temperature, from $\sim 0.5T_c$\ to \tc(see for example
\cite{Klein93,Wagner94}). In this area the surface or edge barriers of
different types can manifest themselves. They prevent the entry and exit of
vortices and consequently provide for the magnetic flux trapping.

To increase pinning the various artificial pinning centers are created in a
superconductor. The most effective among them are defects of columnar
shape. They are generated by irradiating the material in a beam of heavy
ions accelerated up to high energy, $\sim 1$~GeV (see \cite{Civale97} and
references therein). It has been shown in many publications that columnar
defects (CD) have a large pinning potential and enlarge the critical current
density (see for example \cite{Civale97,Kazumata96}). The creation of CD in
the \bi crystals leads to the shift of irreversibility line (IL) towards
high magnetic fields and temperatures (see for example
\cite{Konczykowski95,Zech95}).

However the effect of columnar defects is not always straightforward.
Sometimes the inverse influence of CD on superconducting properties is
observed. An example of such influence has been presented in the work by
J.K.Gregory \textit{et al} \cite{Gregory01}. They have found the decrease
of the penetration field in \bi whiskers after electron and heavy ion
irradiation. The authors had attributed this effect to the suppression of
surface potential barriers due to the CD creation near the superconductor
surface.

The theoretical discussion of the mechanism of the surface barrier
reduction under the CD production was presented by A.E.
Koshelev and V.M. Vinokur \cite{KV01}. In accordance with this discussion,
CD near a surface of superconductor change the interaction of pancake
vortices with a shielding current and with their mirror images. Firstly, CD
serve as obstacles for the shielding current and change its density. Secondly,
additional vortex images are formed behind the column surface. Both
these circumstances make vortex penetration easier and thereby reduce the
surface barrier.

The studies of magnetization of the \bi whiskers irradiated by electron and
Pb-ion beams with comparison with that of pristine samples \cite{Gregory01}
have shown the decrease of both the penetration field and the irreversible
magnetization after irradiation. As in accordance with previous studies
\cite{Aukkar96} the magnetic properties in \bi whiskers are dominated by
surface effects, the penetration field decrease was attributed to the
suppression of surface barriers by CD. But after irradiation the bulk
pinning becomes remarkable and contribute to a sample magnetization, which
was measured in the work \cite{Gregory01}. It is impossible to separate
contributions of surface barrier and bulk pinning in such type of
measurements.

In our study we use another technique, namely, a registration of the
hysteresis loop of microwave absorption (MWA). This method allows one to
separate contributions of bulk pinning (BP) and of surface pinning (SP) by
the shape of loop. The present study of the irreversibility line in the \bi
thin films irradiated with heavy ions confirms the suggestion of
Ref.\cite{Gregory01} and the theoretical estimates \cite{KV01} of
suppression of surface barrier by columnar defects. The results of
measurements of hysteretic microwave absorption have shown that in the
films irradiated with \ur -ions the high temperature part of irreversibility line, which is 
due to surface barrier, shifts towards low temperature and magnetic field.

\section{Experimental}

The \bi films were prepared on a \laal substrate with sputtering technique
as described in Ref.\cite{Wagner93}. They are of 200~nm thickness and have
the superconducting transition temperature near 85~K. They were irradiated
with \ur\ ions in the accelerator facility of the Gesellschaft f\"ur
Schwerionenforschung in Darmstadt. The irradiation doses, expressed in the
dose-equivalent matching field \bfi, were varied from 0.3 to 3~T. The
superconducting transition shifts towards low temperature and broadens
after irradiation of the films. The shift increases with the irradiation dose.
The transition temperature of the film irradiated with dose $B_{\Phi}=2$\,T
equals 78\,K.

We studied irreversible properties of films by measuring the hysteresis of
nonresonant microwave absorption. The detailed description of this method
one can find in Refs.\cite{Shap98,Shap02}. This method is sufficiently
sensitive to enable the investigation of magnetic irreversibility of very
thin superconducting films. Moreover it allows one to separate
contributions of bulk pinning and of surface pinning by the shape of the
hysteresis loop. The measurements were performed in the high-temperature
area, close to \tc, where the surface pinning contribution became
remarkable.

The ESR-spectrometer BER-418s (Bruker) with the working frequency of
9.4\,GHz (X-band) was used to measure the microwave absorption of
superconducting thin films. A sample was placed into the spectrometer
cavity inside the helium gas-flow cryostat. Its orientation was as follows:
the applied DC field \ha was perpendicular to the film plate (parallel to
c-axis), and the microwave field \hmw was in the film plane. A DC field was
modulated with a frequency of 100\,kHz and an amplitude from 0.1 to 10\,Oe.
To record the MWA hysteresis loop \ha was swept with a velocity of $\sim
20\div40$\,Oe/s from cooling field $H_i$\ to $H_i+\Delta H$ and back
($\Delta H=100\div5000$\,Oe). To avoid the instrumental error of detected
hysteresis the signal of electron spin resonance (ESR) of paramagnetic
substance DPPH (diphenylpicrylhydrazyl) is recorded along with the MWA
hysteresis loop. The points of the irreversibility line \hir were obtained
by registration of the field at which the MWA hysteresis collapsed.

\section{Results and Discussion}

\begin{figure}[H]
\begin{center}
\includegraphics[width= 14 cm]{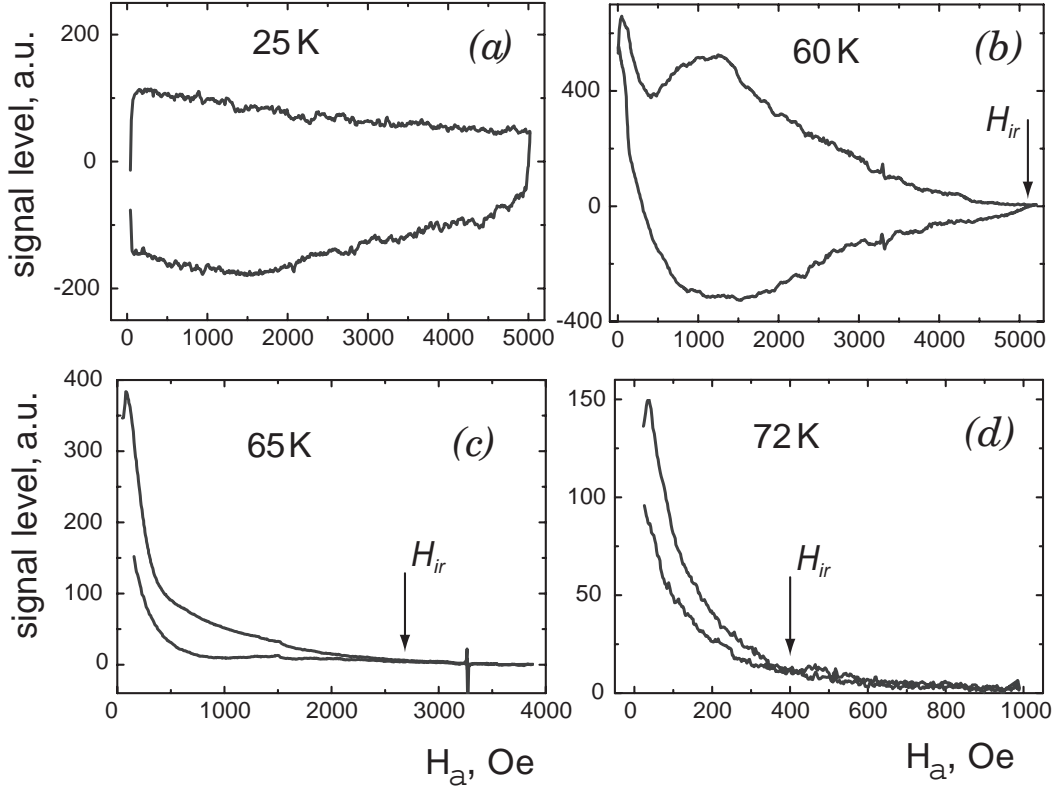}
\end{center}
  \caption{ MWA hysteresis loops obtained on the pristine \bi film \textsf{B1711}
at different temperatures: (a) $T=25$~K, the loop shape is of purely "bulk
pinning type"; (b) $T=60$~K, the small addition of the loop of "surface
pinning type" is seen in low fields, while in high fields "the bulk pinning
loop" predominates and determines \hir ; (c) $T=65$~K, the contributions of BP
and SP to the MWA hysteresis are compatible; (d) $T=72$~K, the hysteresis is
due to surface barrier only and $H_{ir}\simeq400$~Oe. The ESR signal of a 
small piece of DPPH is seen on loops $H_a\simeq3300$~Oe in the panels (b) and (c).  }
  \label{filoops}
\end{figure}

Previously we had found that both the shape of the MWA hysteresis loop and
its dependence on modulation amplitude were governed by the vortex matter
state and by the type of pinning: bulk or surface \cite{Shap02}. One can
exploit these properties of the MWA loop to locate the phase diagram areas
with predominance of bulk pinning or surface barrier. In Fig.\ref{filoops}
the MWA hysteresis loops obtained on the pristine \bi film \textsf{B1711}
at different temperatures are shown. At low temperatures ($T\leq50$~K) the
loop shape corresponds to bulk pinning only \cite{Shap98,Shap02}. At high
temperatures, close to \tc, (Fig.\ref{filoops}d) the loop has the shape,
which is due to surface barrier \cite{Shap02}. Here the point of
irreversibility line ($H_{ir}(72~K)\simeq400$~Oe) is determined by surface
pinning. At intermediate temperatures the mixture of these two types of the
hysteresis is observed (Fig.\ref{filoops}b). At $T=60$~K the small addition
of "surface loop" to the main loop, which is due to bulk pinning, occurs in
low field. The hysteresis disappears at $H_a\simeq 5000$~Oe, and the
boundary of irreversibility is determined by the bulk pinning. At $T=65$~K
(Fig.\ref{filoops}c) the contributions of bulk pinning and surface one are
approximately equal to each other.

\restylefloat{figure}

\begin{figure}[H]
\begin{center}
\includegraphics[width= 8 cm]{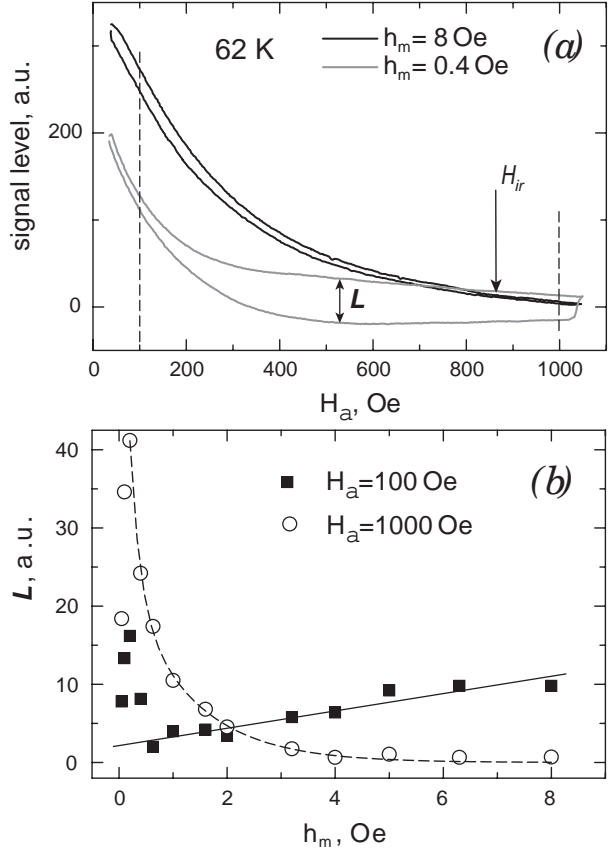}
\end{center}
  \caption{(a) The shape of the MWA hysteresis loop in the film \textsf{B1712}
  irradiated with the dose $B_{\Phi}=0.3$~T at two magnitudes of modulation
  amplitude, 0.4~Oe and 8~Oe, $T=62$~K. Dashed vertical lines indicate fields,
  at which the dependence $L(h_m)$, shown below, is measured. The arrow points
  out the field at which the hysteresis disappears. (b) The dependence of
  hysteresis value on modulation amplitude obtained at two magnitudes of
  applied field, ($\blacksquare$) -- 100~Oe and ($\bigcirc$) -- 1000~Oe. Solid
  straight line and dashed exponential decay are drawn to illustrate the
  different character of the dependence in two cases discussed. }
  \label{fiMix}
\end{figure}

When both types of pinning contribute to a hysteresis we use the difference of their dependence on the field modulation amplitude in order to separate one contribution from another. In Fig.\ref{fiMix}a it is shown the change of a loop shape with increasing the
modulation amplitude \hm from 0.4 to 8~Oe. At small \hm  SP contributes
remarkably in the low field range ($H_a<200$~Oe), and BP predominates in
higher fields. At $h_m=8$~Oe the BP contribution is fully eliminated. A
shape of hysteresis loop is of "pure surface type" over whole field range
from 0 to $H_{ir}\simeq900$~Oe. So this field corresponds to the boundary
of existing the surface barrier, and the irreversibility field has to be
denoted as $H_{ir}^{SP}$, that is due to surface pinning.

The lower part of Fig.\ref{fiMix} shows the dependence of the hysteresis
magnitude $L$\ on the modulation amplitude \hm for two values of applied
field: a low field $H_a=100$~Oe, where the SP contribution is
considerable at any \hm, and a higher field $H_a=1000$~Oe, where the BP
prevails at small \hm. It is seen that at $h_m>0.3$~Oe the surface
contribution increases almost linearly. In order to emphasize this fact the
straight line is drawn along the points of $H_a=100$~Oe. However, the bulk pinning
contribution drastically reduces with increasing modulation amplitude. The curve of exponential decay is added for comparison. Thus if one applies the modulation amplitude
$h_m\sim10$~Oe, it is possible to exclude the bulk pinning contribution and
to investigate the dependence of $H_{ir}^{SP}$\ on temperature and the
irradiation dose.

\begin{figure}[H]
\begin{center}
\includegraphics[width= 10 cm]{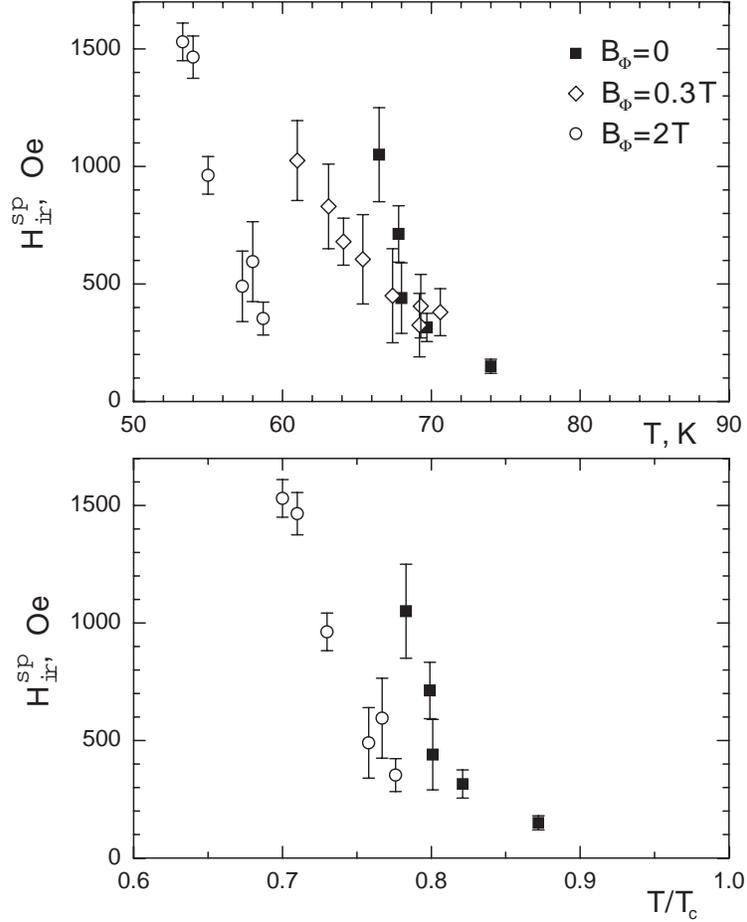}
\end{center}
  \caption{(a) The temperature dependence of the irreversibility field due to
  surface pinning $H_{ir}^{SP}(T)$\ for three samples: ($\blacksquare$) --
  pristine film, ($\lozenge$) -- film irradiated with small dose ($B_{\Phi}=0.3$~T),
  ($\bigcirc$) -- film irradiated with large dose ($B_{\Phi}=2$~T).
  (b) The dependence of the irreversibility field on reduced temperature for
  the pristine film ($\blacksquare$) and film irradiated with dose
  $B_{\Phi}=2$~T  ($\bigcirc$).}
  \label{IL}
\end{figure}

The points of irreversibility line $H_{ir}^{SP}(T)$\ are shown in
Fig.\ref{IL}a for three films, one is pristine and two films ion-irradiated
with various doses, 0.3~T and 2~T. It is clearly seen that IL shifts toward
lower magnetic fields after irradiation. The effect is more pronounced in
the film with large irradiation dose.

To take into account the shift of the superconducting transition
temperature \tc after the ion irradiation the IL points of two samples are
plotted versus reduced temperature $T/T_c$\ in Fig.\ref{IL}b. The shift of the 
IL of irradiated film looks not so considerable with this axis as in
Fig.\ref{IL}a, but it is unambiguous. 

The shift of the "surface" irreversibility line towards the area of low temperatures and magnetic fields indicates that the surface barrier is weakened upon the creation of columnar defects under the heavy-ion irradiation. This fact is in qualitative agreement with the theoretical prediction \cite{KV01}. Unfortunately, it is impossible to make quantitative comparison of our experimental data with the theoretical calculation  since only the changes of the penetration field were estimated in paper \cite{KV01}, but not the irreversibility field. 
To estimate the decrease of the surface potential barrier under the CD generation quantitatively the comparison of the theoretically calculated IL shift with the experimental results is needed. This might be a subject of future publication.

\section{Conclusion}

Thus, using the method of hysteretic microwave absorption measurements and
its dependence on the applied magnetic field modulation we could isolate
the contribution of surface pinning and watch the changes of the
irreversibility line with increasing the dose of irradiation of the \bi
thin films with uranium ions. In agreement with the theoretical prediction
\cite{KV01} the suppression of surface barrier by columnar defects has been
found. It manifests itself as the shift of irreversibility line towards low
temperature and magnetic field. The effect is most significant for the
films irradiated with large doses, $B_{\Phi}\geq1$\,T. \bigskip\bigskip


\section*{Acknowledgements}

This work was performed in the frame of Joint Program for scientific collaboration between Russian Academy of Science and Deutsche Forschungsgemeinschaft. The work was partially  supported by the Russian Ministry of Industry and Science under State Contract No. 40.012.1.1.1356, Russian Foundation for Basic Research (grant No. 03-02-96230) and the NIOKR Fund of the Academy of Sciences of Tatarstan through grant No. 06-6.2-234 (Yu.T.).



\end{document}